\documentclass[a4paper]{amsart}

\usepackage[latin1]{inputenc}
\usepackage{mysymb}
\usepackage{myref}

\numberwithin{equation}{section}

\begin{document}
\title{Note on mirror symmetry and coisotropic D-branes on tori}
\author{Christian van Enckevort}
\address{Johannes Gutenberg Universität\\Mainz\\Germany}
\email{enckevor@mathematik.uni-mainz.de}
\maketitle

\begin{abstract}
We describe mirror symmetry on higher dimensional tori, paying special
attention to the behaviour of D-branes under mirror symmetry. To find the
mirror D-branes the description of mirror symmetry on D-branes due to
Ooguri, Oz en Yin is used. This method allows us to deal with the
coisotropic D-branes recently introduced by Kapustin and Orlov. We
compare this to the description of mirror symmetry on D-branes using
the Fourier-Mukai transform of charges.
\end{abstract}

\section{Introduction}
Since its first breakthrough in 1991 (see \cite{COGP91}) mirror
symmetry has been developed in many directions. This topic has
stimulated a lively exchange of ideas between mathematics and physics.
Since the formulation of the homological mirror symmetry conjecture by
Kontsevich in \cite{Kon94} and the introduction of D-branes by
Polchinsky (see \cite{Pol95,Pol96}), D-branes have started to play an
increasingly important role. We will restrict our attention to
BPS-branes on Calabi-Yau spaces. These come in two classes the
(special) Lagrangian D-branes and the holomorphic D-branes. These are
also often called the A-branes and B-branes, because the Lagrangian
branes can also be discussed in the topological A-model, whereas the
B-branes can be included into the topological B-model.

Kontsevich proposes to combine each of these classes of D-branes into
a category. The holomorphic D-branes should lead to the category which
is known among mathematicians as the derived category of coherent
sheaves. The Lagrangian D-branes should give rise to the so-called
Fukaya category (or better a derived version thereof). The homological
mirror symmetry conjecture then says that for a mirror pair of
Calabi-Yau manifolds $X$ and $Y$ the derived Fukaya category of $X$
should be equivalent to the derived category of coherent sheaves of
$Y$.

Recently these D-brane categories have been investigated more closely
by physicists (see e.g. \cite{Dou00} and references therein). These
investigations led to interesting conjectures and results about
stability, the relation to Witten's open string field theory etc.
Another interesting line of research was initiated by Kapustin and
Orlov in \cite{KO01}, where it was shown that the description of
A-branes as Lagrangian branes is in general incomplete and extra
objects corresponding to so-called coisotropic manifolds have to be
introduced. To find these objects Kapustin and Orlov used the
description of mirror symmetry introduced in \cite{OOY96}.

In this article I want to give a more concrete description of these
extra objects. In \cite{KO01} they were studied from a general
perspective. Here we will restrict ourselves to even-dimensional tori
as a special example of Calabi-Yau manifolds.  This restriction allows
us to give a much more concrete description of these still rather
mysterious objects. We also investigate some general properties of
these objects and check that they match the expectations from mirror
symmetry.

\section{Mirror symmetry on tori}
\subsection{Notation}
Let us first briefly discuss the notation that we will use throughout
this text. We write $X = T^n \times T^n$ for a $2n$-dimensional torus
and $Y = T^n \times \check{T}^n$ for its mirror. We will use
coordinates $(x, y) \in \R^n/\Z^n \times \R^n/\Z^n$ on $X$ and $(x,
\check{y}) \in \R^n/\Z^n \times \R^n/\Z^n$ on $Y$. The data describing
objects on $X$ will have a tilde, to distinguish them from the data
describing objects on $Y$. The SYZ-fibration is given by projection to
the $x$-coordinates. We will discuss the A-model on $X$ and the
B-model on $Y$.

The tangent bundle is trivial and can be written as $TX = X \times
\R^n \times \R^n$. A Ricci flat metric can then be identified with a
positive definite $2n \times 2n$-matrix $g$, the complex structure
with a $2n \times 2n$-matrix squaring to $-I_{2n}$. The compatibility
of the complex structure and the metric can be written as $J^t g J =
g$. It follows that $\omega = g J$ is an antisymmetric matrix defining
the Kähler form on $X$.

The complex structure can also be defined by requiring $z := \tau x +
y$ to define complex coordinates on $X$. Here $\tau = \tau_1 + i
\tau_2$ is a complex $n \times n$-matrix. For simplicity we will
assume that $\tau_1 = 0$ and that $\tau_2$ be symmetric. This mirror
symmetric to requiring the B-field to vanish. The relation between
these two ways of define the complex structure is that in term of
$\tau$ the matrix $J$ can be written as
\begin{equation}
\label{eq:cplxstr}
  J = \begin{pmatrix}
          0       & \tau_2^{-1} \\
          -\tau_2 & 0
      \end{pmatrix}.
\end{equation}
On a Calabi-Yau manifold we also have a holomorphic $(n,0)$-form
$\Omega$. On a torus this form can be written as $\Omega = \dop^n z$.

\subsection{Gluing matrices}
We will use the description of D-branes in Calabi-Yau spaces that was
given in \cite{OOY96}. This description was also used by Kapustin and
Orlov to give a general description of the non-Lagrangian A-branes.
In \cite{OOY96} everything is discussed in terms of local coordinates
and it is not completely clear how to formulate all of it in a
coordinate independent way, especially because in general the mirror
manifold and the original manifold are really different manifolds.
Fortunately the mirror manifold of a torus is a torus again and in
addition we can use global flat coordinates on the tori (as discussed
above). So we give a description tailored for tori and ignore
subtleties that do not matter for tori.

Let $X$ be a complex torus as discussed above. The worldsheet $\Sigma$
is a two-dimensional surface with metric $h$. The fields\footnote{I use
  the notation from \cite{Enc00}. The fields $x$, $\psi_+$ and
  $\psi_-$ correspond to the fields $X$, $\psi_L$ and $\psi_R$ in
  \cite{OOY96}.} of the sigma model are the bosonic field $x$ which is
a map $\Sigma \rightarrow X$ and the fermionic fields $\psi_+$ and
$\psi_-$, which are sections of $x^*(TX) \otimes K^{1/2}$ and $x^*(TX)
\otimes \bar{K}^{1/2}$ respectively. Here $K$ is the bundle of
$(1,0)$-forms on the surface $\Sigma$.

Let $S$ be a submanifold of $X$. We will assume that $S$ corresponds
to an affine subspace of $\R^n \times \R^n$ (affine D-brane). This
assumption will allow us to describe the D-brane in terms of a matrix.
We want to discuss boundary conditions for open strings ending on $S$.
For string theory on a Calabi-Yau manifolds there are two special
classes of boundary conditions that preserve half of the
supersymmetry, which are called A- and B-type boundary conditions. For
affine D-branes on a torus they can be described using an $2n \times
2n$-matrix $R$. On the boundary of $\Sigma$ the following boundary
conditions should hold
\[
  \partial x^\mu = R^\mu{}_\nu \bar{\partial} x^\nu, \quad
  \psi^\mu_+ = \pm R^\mu{}_\nu \psi^\nu_-.
\]
The matrix $R$ should be orthogonal with respect to the metric $g$ on
the Calabi-Yau manifold $X$. As explained in \cite{OOY96} the
codimension of the D-brane with gluing matrix $R$ is given by the
dimension of the eigenspace $E_{-1}$ for eigenvalue $-1$ of $R$. The
orthogonal complement $E_{-1}^\perp$ of that eigenspace is the tangent
space to the D-brane. The restriction of $R$ to that eigenspace can be
written as 
\[
  R|_{E_{-1}^\perp} = (\tilde{g}+F)^{-1}(\tilde{g}-F).
\]
Here $F$ is an antisymmetric matrix defining the curvature of the
connection on the D-brane and $\tilde{g}$ is the restriction of the
metric $g$ to $E_{-1}^\perp$.

Whether these boundary conditions are of A- or B-type depends on the
additional conditions that we impose on $R$. The conditions for A-type
boundary conditions can be formulated as
\[
\begin{split}
  R^t \omega R &= -\omega, \\
     R^*\Omega &= \bar{\Omega}.
\end{split}
\]
For B-type boundary conditions $R$ should satisfy
\[
\begin{split}
  R^t \omega R &= \omega, \\
     R^*\Omega &= e^{i\theta} \Omega,
\end{split}
\]
for some $\theta \in \R$.

\subsection{T-duality and mirror symmetry}
When the target manifold $X$ has a fibration with a torus as fibre,
one can apply T-duality and replace the torus by its dual. Because
T-duality affects the boundary conditions, we should describe what
T-duality does with the gluing matrices discussed above. Following
\cite{OOY96} we will describe T-duality using a matrix $T$. This
matrix should square to $I_{2n}$, be symmetric (or equivalently
orthogonal) with respect to the metric $g$ and satisfy $T J T = -J$.
In terms of this matrix mirror symmetry on D-branes should be
described as follows. If we start with a D-brane with gluing matrix $R$,
then mirror D-brane should have gluing matrix $R' = RT$. However, for
this description to be valid there must be a relation between the
choice of coordinates on $X$ and on $Y$ (because $R$, $T$ depend on
the coordinates on $X$ and $R'$ on the coordinates on $Y$). We will
suppose this description is valid for coordinates on $X$ such that
$g=I_{2n}$ and $J=\left ( \begin{smallmatrix} 0 & I_n \\ -I_n &
    0\end{smallmatrix} \right )$. We will use appropriate coordinate
transformations to transform to this situation. To see how that works,
we must have a closer look at the background fields. 

In the coordinates that we are using here $T = \diag(I_n, -I_n)$. For
this matrix to be symmetric with respect to the metric $g$, we have to
require that $g$ has the following block diagonal form $g =
\diag(g_{xx}, g_{yy})$, where $g_{xx}$ and $g_{yy}$ are positive
definite symmetric $n \times n$-matrices. Recall that we consider a
complex structure of the block form \pref{eq:cplxstr}. Note that $J$
automatically satisfies $T J T = -J$. The condition that $J$ be
orthogonal with respect to $g$ yields $g_{xx} = \tau_2^t g_{yy}
\tau_2$. Let $S_{xx}$ be an $n \times n$-matrix such $g_{xx} =
S_{xx}^t S_{xx}$. Then we can introduce new coordinates $(x',y')$
defined by the following equation
\[
  \begin{pmatrix} x' \\ y' \end{pmatrix} =
  S \begin{pmatrix} x \\ y \end{pmatrix} =
  \begin{pmatrix}
    S_{xx} & 0 \\
    0      & S_{xx} \tau_2^{-1}
  \end{pmatrix}
  \begin{pmatrix} x \\ y \end{pmatrix}.
\]
With respect to these new coordinates the metric and the complex
structure have the standard form discussed above. In the new
coordinates the D-brane is described by the gluing matrix $\tilde{R} := S
R S^{-1}$. Note that $T$ is invariant under such coordinate
transformations. So the mirror D-brane is described by $\tilde{R}' =
\tilde{R} T$. However, this description is in terms of certain unknown
coordinates $(x'',\check{y}')$ on $Y$.  Recalling that $\check{y}$ is
dual to $y$, a reasonable guess for the definition of these new
coordinates is
\[
  \begin{pmatrix} x'' \\ \check{y}' \end{pmatrix} =
  \check{S} \begin{pmatrix} x \\ \check{y} \end{pmatrix} =
  \begin{pmatrix}
    S_{xx} & 0 \\
    0      & S_{xx}^{-t} \tau_2^t
  \end{pmatrix}
  \begin{pmatrix} x \\ \check{y} \end{pmatrix}.
\]
In the coordinates $(x'', \check{y}') = (x', \check{y}')$, the metric
on $Y$ is the standard metric, so in the coordinates $(x, \check{y})$
the metric is
\[
  \check{g} = \check{S}^t \check{S} =
  \begin{pmatrix}
    g_{xx} & 0 \\
    0      & g_{yy}^{-1}
  \end{pmatrix}.
\]
Similarly we find the complex structure
\[
 \check{J} = \check{S}^{-1}
   \begin{pmatrix}
      0  & I_n \\
    -I_n &  0
   \end{pmatrix}
   \check{S} =
  \begin{pmatrix}
    0 & g_{xx}^{-1} \tau_2^t \\
    -\tau_2^{-t} g_{xx} & 0
  \end{pmatrix}.
\]
Comparing to \pref{eq:cplxstr}, we see $\check{\tau}_2 = \tau_2^{-t}
g_{xx}$. Together with $\check{g}_{xx} = g_{xx}$ and
$\check{g}_{\check{y}\check{y}} = g_{yy}^{-1}$, this defines the
mirror map on the background fields. To compare this with the mirror
map in e.g. \cite{Enc00}, let us define $k = g_{yy} \tau_2 =
\tau_2^{-t} g_{xx}$.  One can easily check that the Kähler form can be
written as an antisymmetric block matrix
\[
  \begin{pmatrix}
     0 & k^t \\
    -k & 0
  \end{pmatrix}.
\]
So $k$ parametrises the Kähler structure. In this notation the mirror
map on the background fields is given by $\check{k} = \tau_2$ and
$\check{\tau}_2 = k$, which matches the result in \cite{Enc00}.


\subsection{D-brane charges}
To D-branes one can associate so-called charges. These are easiest to
define for B-branes. For a vector bundle $E$ on a torus $Y$ the charge
vector is given by the Chern class $\ch(E)$. For more general
Calabi-Yau manifolds one usually uses the so-called Mukai vector $v(E)
:= \ch(E)\sqrt{\td(Y)}$. Of course for a torus $\td(Y) = 1$, so $v(E)
= \ch(E)$. Using these charges, we can define the Euler characteristic
of a pair $(E,F)$ of B-branes
\[
  \chi(E,F) := \int_Y \ch(F) \cup \ch(\check{E}) \cup \td(Y)
    = \int_Y v(F) \cup v(\check{E}).
\]
According to the Riemann-Roch theorem we have
\begin{equation}
\label{eq:RR}
  \chi(E,F) = \sum_i (-1)^i h^i(E,F),
\end{equation}
where $h^i(E,F) := \dim \Ext^i(E,F)$. For the purposes of this article
these definitions suffice. However, it has been suggested in the
literature that one should really use K-theory instead of cohomology
classes.

For A-branes the definition is quite simple in the case of Lagrangian
branes. In this case the charge vector is given by $r \PD(L)$, where
$\PD(L)$ is the Poincaré dual of the Lagrangian submanifold $L$ and
$r$ is the rank of the flat bundle on $L$. For more general
coisotropic submanifolds $Z \subset X$ with a bundle $F$ defined on
$Z$, we can define the charge as follows
\[
  c(Z,F) := i_*(\ch(F)),
\]
where $i : Z \rightarrow X$ is the embedding.  For Lagrangian
submanifolds this reproduces the definition formulated above. The
natural guess for the definition of the Euler characteristic for a
pair of A-branes $(Z_i, F_i)$ ($i=1,2$) is
\[
  \chi((Z_1,F_1),(Z_2,F_2)) := \int_X c(Z_2,F_2) \cup 
    c(Z_1,\check{F}_1).
\]
Here $\check{F}_1$ is the dual bundle of $F_1$ on $Z_1$. In the
simplest case of two Lagrangian submanifolds $L_i$ with rank $r_i$
flat vector bundles $F_i$, this Euler characteristic reduces to
\[
  \chi((L_1,F_1),(L_2,F_2)) := r_1 r_2 \int_X \PD(L_2) \cup \PD(L_1)
  = r_1 r_2 L_1 \cdot L_2,
\]
where $L_1 \cdot L_2$ counts the intersection points of $L_1$ and
$L_2$ with sign. Recall that a basis of the complex $\Hom^*((L_1,F_1),
(L_2,F_2))$ in the Fukaya category is given by the intersection points
(tensored with linear maps between the corresponding fibres of $F_1$
and $F_2$). Because the Euler characteristic of a complex is equal to
the Euler characteristic of its cohomology, the analog of \pref{eq:RR}
holds in this simple case. It is unclear if a similar result is valid
for coisotropic branes, because for that case we do not have a
definition of the space of morphisms.

The mirror map on charge vectors is expected to be given by fibrewise
Fourier-Mukai transform $\FM: A^*_Y \rightarrow A^*_X$ of differential
forms, where $A^*_X$ denotes the space of differential forms on $X$
and similarly for $A^*_Y$. This map is defined as follows
\[
  \FM(\alpha) = \int_{T^n_{\check{y}}} \alpha e^{-\langle \dop
  \check{y}, \dop y \rangle}.
\]
Let $\mu$ be the mirror map of D-branes, mapping holomorphic D-branes
on $Y$ to objects of the Fukaya category on $X$. Then one expects
\[
  \FM(\ch(E)) = c(\mu(E)).
\]
The charges depend on the support of the D-brane and the curvature of
the connection. This is also the information contained in the
gluing matrix. In both cases we have a description of how mirror symmetry
should act. In the sequel we will analyse mirror symmetry using both
descriptions and compare the results.

\section{Mirror symmetry for D-branes}
\subsection{Line bundles}
Let us first consider a line bundle $L$ on $Y$ and its mirror object on
$X$. The first Chern class of this line bundle can be written as
\[
  c_1(L) = \tfrac{1}{2} \langle \dop x, A \dop x \rangle +
  \tfrac{1}{2} \langle \dop \check{y}, B \dop \check{y} \rangle +
  \langle C \dop x, \dop \check{y} \rangle.
\]
Because $c_1(L)$ should be integral, we see that the entries of the $n
\times n$ matrices $A$, $B$ and $C$ have to be integers. Note that the
matrices $A$ and $B$ can be chosen to be antisymmetric. To make
contact with the usual notation in physics, the curvature matrix $F$
of the line bundle $L$ can, in terms of the matrices $A$, $B$ and
$C$, be written as $F = \left ( \begin{smallmatrix} A & C^t \\ -C & B
\end{smallmatrix} \right )$.

Another requirement is that $c_1(L)$ should be a $(1,1)$-form.  Recall
that complex coordinates on $Y$ are defined by $z = \tau x + y$, where
for simplicity we assume that $\tau=i\tau_2$ is purely imaginary.  A
simple computation shows that $y = (z + \bar{z})/2$ and $x =
(2i\tau_2)^{-1} (z - \bar{z})$. So the $(0,2)$-part of $c_1(L)$ is
given by
\[
  c_1(L)^{(0,2)} = -\tfrac{1}{8} \langle \dop \bar{z}, \tau_2^{-t} A
  \tau_2^{-1} \dop \bar{z} \rangle + 
  \tfrac{1}{8} \langle \dop \bar{z}, B \dop \bar{z} \rangle -
  \tfrac{i}{4} \langle \dop \bar{z}, \tau_2^{-t} C \dop \bar{z} \rangle.
\]
So for $c_1(L)$ to be a $(1,1)$-form $\tau_2^{-t} C$ should be
symmetric and the antisymmetric matrices $\tau_2^{-t} A
\tau_2^{-1}$ and $B$ should be equal.
  
For the SYZ-fibration that we fixed above, mirror symmetry is given by
the T-duality matrix $T = \left ( \begin{smallmatrix} I_n & 0 \\ 0 &
    -I_n \end{smallmatrix} \right )$. As discussed above a very
general description of T-duality can be given in terms of the
gluing matrices. For simplicity we will first assume that the metric is
the standard metric given by the matrix $g = I_{2n}$. In that case the
gluing matrix for the line bundle $L$ is given by $R =
(I_{2n}+F)^{-1}(I_{2n}-F)$. The gluing matrix of the dual A-brane is then
given by $\tilde{R} = RT$. For a geometric interpretation of the
B-brane in terms of a Lagrangian submanifold we need that $\tilde{R}$
is a symmetric matrix. Using the antisymmetry of $F$ and the symmetry
of $T$ this condition can be written as $(I_{2n}-F)T(I_{2n}-F) =
(I_{2n}+F)T(I_{2n}+F)$. A small calculation with block matrices shows
that condition is met when $A$ and $B$ vanish. So we find that the
geometric case is exactly the case investigated in \cite{Enc00}.

In this case $\tilde{R}$ can easily be calculated. We can write
$\tilde{R} = \bigl ( \begin{smallmatrix} \alpha & \gamma \\ \gamma^t &
  \beta \end{smallmatrix} \bigr )$. If we write the defining equation
for $\tilde{R}$ as $(I_{2n}+F) \tilde{R} = (I_{2n}-F) T$, it is a
matter of some careful computations with block matrices to find
\[
  \tilde{R} = 
     \begin{pmatrix}
       I_n - 2C^t(I_n + CC^t)^{-1} C & 2C^t(I_n + C C^t)^{-1} \\
       2(I_n + CC^t)^{-1} C          & -I_n + 2CC^t(I_n + CC^t)^{-1}
     \end{pmatrix}
\]
One can verify that the eigenspace with eigenvalue $1$ is $\{ (
 \begin{smallmatrix} x \\ C x \end{smallmatrix} ) \}$ and
that its orthogonal complement $\bigl \{ \bigl (
\begin{smallmatrix} -C^t y \\ y \end{smallmatrix} \bigr ) \bigr \}$ is the
eigenspace with eigenvalue $-1$. This is in complete agreement with
\cite{Enc00}, where the mirror brane was found to be given by the
equation $y = C x + \alpha$ for some shift vector $\alpha \in \R^n$,
which we cannot find in this way.

It is interesting to check what this looks like when the metric is not
the standard metric. As we did above, we will use a coordinate
transform given by a block diagonal matrix $S = \diag(S_{xx}, S_{xx}
\tau_2^{-1})$ to transform to the standard metric and the standard
complex structure (note however, that we are starting on $Y$ and
transforming to $X$ instead of the other way around). We will denote
the matrices on the B-side with respect to the new orthogonal
coordinates with primes. We find $R' = (I_{2n} + F')^{-1} (I_{2n}
-F')$. Again $R'$ will only be symmetric when $F'= S^{-t} F
S^{-1}$ has the form $F' = \bigl ( \begin{smallmatrix}0 & {C'}^t \\
    -C' & 0
  \end{smallmatrix} \bigr )$, where $C' = S_{xx}^{-t} \tau_2^t
C S_{xx}^{-1}$. Because $T' = T$, we can use the result above to find
$\tilde{R}'$. However, this result is still with respect to
transformed coordinates. The transformation to the original
coordinates on $X$ is given by $\tilde{S}=\diag(S_{xx}, S_{xx}^{-t}
\tau_2^t)$. So the eigenspace of $\tilde{R}$ for the eigenvalue $1$
can be written as
\[
  E^{\tilde{R}}_1 = \bigl \{ \tilde{S}^{-1} \bigl ( \begin{smallmatrix} x'
  \\ C' x' \end{smallmatrix} \bigr ) \mid x' \in \R^n \bigr \} 
  = \bigl \{ \bigl ( \begin{smallmatrix} S_{xx}^{-1} x' \\ 
   C S_{xx}^{-1} x' \end{smallmatrix} \bigr ) \mid x' \in \R^n \bigr \}
  = \{ \bigl ( \begin{smallmatrix} x \\ 
   C x \end{smallmatrix} \bigr ) \mid x \in \R^n \}.
\]
So we see that the mirror brane does not depend on the metric.

The more complicated case is when $\tilde{R}$ is not symmetric. We
will first describe this case in terms of charges. Using an orthogonal
basis transformation we can write obtain a basis compatible with the
decomposition $\R^{2n} = \ker A \oplus \ker B \oplus (\ker A)^\perp
\oplus (\ker B)^\perp$. Note that orthogonal and perpendicular is with
respect to the standard metric on $\R^{2n}$ (and not the one inducing
the metric on the torus). The corresponding coordinates will be
denoted with $(x_0, \check{y}_0, x_1, \check{y}_1)$.

Using these coordinates we can write $A = \left ( \begin{smallmatrix}
    0 & 0 \\ 0 & A' \end{smallmatrix} \right )$ (w.r.t.\ $x_0$,
$x_1$), $B = \left ( \begin{smallmatrix} 0 & 0 \\ 0 & B'
  \end{smallmatrix} \right )$ (w.r.t.\
$\check{y}_0$, $\check{y}_1$) and $C = \left ( \begin{smallmatrix}
    C_{00} & C_{01} \\ C_{10} & C_{11} \end{smallmatrix} \right )$
(w.r.t.\ $\check{y}_0$, $\check{y}_1$ and $x_0$, $x_1$). In this
notation the matrix $F$ can be written as
\[
  F = \begin{pmatrix}
           0    & C_{00}^t &     0   & C_{10}^t \\
        -C_{00} &   0      & -C_{01} &   0      \\
           0    & C_{01}^t &     A'  & C_{11}^t \\
        -C_{10} &   0      & -C_{11} &   B'
      \end{pmatrix}
\]
The charge on the B-side is given by the Chern character 
\[
  \ch(L) = e^{\frac{1}{2} \langle \dop v, F \dop v \rangle},
\]
where $v = (x_0, \check{y}_0, x_1, \check{y}_1)$.  On the A-side the
D-brane is given by a coisotropic submanifold $Z \subset X$. On the
submanifold $Z$ a $C^\infty$ vector bundle $E$ with connection
$\nabla$ is defined. The charge of such a D-brane should be defined as
$i_* \ch(E)$, where $i: Z \rightarrow X$ is the inclusion.  A
fibrewise Fourier-Mukai transform on differential forms should map a
representative of the charge cohomology class on the B-side to one
representing the charge on the A-side, i.e.,
\[
\begin{split}
  \FM(\ch(L)) &= \int_{T^n_{\check{y}}} e^{\frac{1}{2} \langle \dop v, 
  F \dop v \rangle}
  e^{-\langle \dop \check{y}_0, \dop y_0 \rangle - \langle
  \dop \check{y}_1, \dop y_1 \rangle } \\
  &= \int_{T^n_{\check{y}}} e^{\frac{1}{2} \langle \dop x_1, A' \dop
  x_1 \rangle + \frac{1}{2} \langle \dop y_1, B' \dop y_1 \rangle 
  + \langle C_{00} \dop x_0 + C_{01} \dop x_1 - y_0, 
  \dop \check{y}_0 \rangle} \\
  &\quad \times e^{\langle C_{10} \dop x_0 + 
  C_{11} \dop x_1 - \dop y_1, \dop \check{y}_1 \rangle }.
\end{split}
\]
Writing $T^n_{\check{y}}$ as $T^{n-p}_{\check{y}_0} \times T^p_{\check{y}_1}$
and integrating over $T^p_{\check{y}_0}$, we obtain using \pref{eq:deltaform}
\[
\begin{split}
  \FM(\ch(L)) &= e^{\frac{1}{2} \langle \dop x_1, A' \dop x_1 \rangle}
  \dop^p(C_{00} \dop x_0 + C_{01} \dop x_1 - y_0) \\
  &\quad \times \int_{T^{n-p}_{\check{y}_1}} e^{\frac{1}{2} \langle \dop y_1, 
  B' \dop y_1 \rangle + \langle C_{10} \dop x_0 + C_{11} \dop x_1 
  - \dop y_1, \dop \check{y}_1 \rangle}.
\end{split}
\]
Using \pref{eq:complsqr} we find
\begin{multline}
\label{eq:FM}
  \FM(\ch(L)) = 
  \dop^p(C_{00} x_0 + C_{01} x_1 - y_0) \\
  \times e^{\frac{1}{2} \langle \dop x_1, A' \dop x_1 \rangle}
  e^{\frac{1}{2} \langle C_{10} \dop x_0 + 
  C_{11} \dop x_1 - \dop y_1,
  (B')^{-1} (C_{10} \dop x_0 + C_{11} \dop x_1 - \dop y_1)
  \rangle} \sqrt{\det(B')}.
\end{multline}
This is a very nice formula, but the interpretation is a bit
complicated. The easy case is when $A=B=0$, which we already discussed
in terms of gluing matrices above. In that case we only keep $x_0$, $y_0$
and $C_{00}$. Dropping the indices $0$, we obtain the following
formula
\[
  \FM(\ch(L)) = \dop^n(C x -y),
\]
which is the Poincaré dual of the Lagrangian submanifold given by the
equation $y = C x$. So this is in full agreement with previous results.

In the general case we expect this to be of the form $\PD(Z)\ch(F)$,
where $Z \subset X$ is a coisotropic submanifold of $X$ and $F$ is a
vector bundle defined on $Z$. If we define
\[
  Z := \{ (x_0, C_{00} x_0 + C_{01} x_1, x_1, y_1) \mid \text{$x_0 \in
  \R^p$, $x_1,y_1 \in \R^{n-p}$} \},
\]
then the first factor in \pref{eq:FM} can be identified with the
Poincaré dual $\PD(Z)$ of $Z$. The second line of \pref{eq:FM} should
therefore be interpreted as $\ch(F)$. The degree $0$ part of the Chern
character is the rank of the vector bundle. So we see that $F$ has
rank $\sqrt{B'}$. The 2-form 
\[
\begin{split}
  & \tfrac{1}{2} \langle \dop x_0, C_{10}^t (B')^{-1} 
  C_{10} \dop x_0 \rangle +
  \tfrac{1}{2} \langle \dop x_1, (A' + C_{11}^t (B')^{-1} 
  C_{11}) \dop x_1 \rangle \\
  & + \tfrac{1}{2} \langle \dop y_1,
  (B')^{-1} \dop y_1 \rangle -
  \langle \dop x_0, C_{10}^t \dop y_1 \rangle - \langle \dop
  x_1, C_{11}^t (B')^{-1} \dop y_1 \rangle +
  \langle \dop x_0, C_{10}^t C_{11} \dop x_1 \rangle
\end{split}
\]
is not integral on $\Z^p \times \Z^p \times \Z^{n-p} \times
\Z^{n-p}$. However, there exists an integral $\frac{1}{2} (n-p) \times
\frac{1}{2}(n-p)$-matrix $B''$ such that
\[
  B' = \begin{pmatrix} 
             0    & B'' \\
         -(B'')^t &  0
       \end{pmatrix}.
\]
Using this matrix we can define the lattice $\Lambda_{B''} := \Z^p
\times \Z^p \times \Z^{n-p} \times B''\Z^{\frac{1}{2}(n-p)} \times
\Z^{\frac{1}{2}(n-p)}$. The above 2-form is integral on this lattice.
So we can find a line bundle $L$ on the torus $\tilde{Y} =
\R^{2n}/\Lambda_{B''}$ with first Chern class given by the above
2-form. Using the isogeny $i: \tilde{Y} \rightarrow Y$, we can define
the vector bundle $\tilde{F} := i_*L$. The restriction of $\tilde{F}$
to $Z$ is then the vector bundle on $Z$ that we are looking for.

Checking if this matches the description in terms of gluing matrices is
rather complicated. In the coordinates discussed above mirror symmetry
is given by the matrix $T=\diag(I_p, -I_p, I_{n-p}, -I_{n-p})$. The
gluing matrix $\tilde{R}$ of the mirror D-brane is given by the equation
$(I_{2n} + F) \tilde{R} = (I_{2n} - F) T$. Recall that the
gluing matrices are orthogonal matrices, so the eigenspaces for different
eigenvalues are orthogonal. It is again easiest to use a suitable
coordinate transform so that we only have to deal with the standard
metric and the standard complex structure. Then one can check that the
orthogonal complement of the eigenspace for eigenvalue $-1$ of
$\tilde{R}$ is indeed given by $Z$. Further checks get increasingly
complicated.

\subsection{Higher rank bundles}
To describe higher rank bundles on the B-side and their mirrors on the
A-side, we use the classification of semi-homogeneous vector bundles
on tori (see Chapter~4 in \cite{Enc00}). Such vector bundles can be
constructed out of line bundles on tori using isogenies and tensor
product with flat line bundles. We will discuss these two
possibilities in turn.

\subsubsection{Tensor product with flat vector bundle}
The simplest possibility is tensor product with a flat vector bundle.
Flat vector bundles correspond to representations of the fundamental
group. For the torus $T^n \times \check{T}^n$ on the B-side, we have
$\pi_1(T^n \times \check{T}^n) = \Z^n \times \Z^n$. We can always use
a holomorphic bundle isomorphism to find an equivalent bundle $F$ such
that the representation is trivial on the second $\Z^n$-factor. Such a
representation also induces a vector bundle on the base torus $T^n$.
This bundle on the base torus can be pulled back to a vector bundle
$\tilde{F}$ on the mirror torus $T^n \times T^n$. One has
\[
  \mu(E \otimes F) = \mu(E) \otimes \tilde{F}.
\]

\subsubsection{Isogenies}
Another way to construct vector bundles of rank greater than one, is
to use isogenies. For a general isogeny it is not clear what the
effect on the mirror will be, but there are two special cases which
can be described. These two cases are when the isogeny is compatible
with the SYZ-fibration.

The easiest of these two special cases is an isogeny $i: \tilde{T}^n
\rightarrow T^n$ on the base torus
alone. This simply commutes with the mirror map, so we have
\[
  \mu(i_*(L)) = i_*(\mu(L)).
\]
Here we slightly abuse notation by using $i$ for both isogenies
$\tilde{T}^n \times T^n \rightarrow T^n \times T^n$ and $\tilde{T}^n
\times \check{T}^n \rightarrow T^n \times \check{T}^n$.

The second case is an isogeny on the fibres alone. Let $i: \tilde{T}^n
\rightarrow T^n$ be an isogeny. This induces an isogeny
$\check{\imath}: \check{T}^n \rightarrow \check{\tilde{T}}^n$. Let us
also write $i$ and $\check{\imath}$ for the corresponding maps $T^n
\times \tilde{T}^n \rightarrow T^n \times T^n$ and $T^n \times
\check{T}^n \rightarrow T^n \times \check{\tilde{T}}^n$, that are the
identity on the first factor. The mirror map intertwines
$\check{\imath}^*$ and $i_*$ as follows
\[
  \mu(\check{\imath}^*(L)) = i_*(\mu(L)).
\]
The isogeny that we used above, can be interpreted in this way.
Similarly, the mirror map should also intertwine $\check{\imath}_*$
and $i^*$.


\section{Conclusions and outlook}
We have given a preliminary description of the mirror objects of
semi-homogeneous vector bundles. These include D-branes that are
coisotropic, but not Lagrangian. We have also investigated some of
their properties and the results fit in nicely with the expectations
from physics in particular from mirror symmetry. Semi-homogeneous
vector bundles are conjectured to generate in a certain sense the full
derived category. So we may hope that this provides us with a fairly
complete understanding of the mirror map on the level of objects.

To do these calculations we restricted the class of background fields
that we considered. In \cite{Enc01b} we will discuss an extension to
more general background fields. This should also clarify the role of
the B-field. In that article we will also give some more details on
things that were left rather sketchy in this article.

Another restriction is that we described the mirror map only on
objects.  However, it should be a functor of D-brane categories, so it
should also define a map on morphisms. Here we have made less
progress, because it remains difficult to define morphisms in the
Fukaya category, especially when non-Lagrangian objects are involved.

\appendix
\section{Gaussian integrals with differential forms on a torus}
Let $T^{2m}$ be an even dimensional torus and let $A$ be a
nondegenerate antisymmetric $2m \times 2m$-matrix. We can choose
coordinates $y_1, \dots, y_{2m}$, such that $A = \left (
  \begin{smallmatrix} 0 & B \\ -B^t & 0 \end{smallmatrix} \right
)$. This allows us to write the Gaussian integral, that we want to
calculate, as follows
\[
  \int_{T^{2m}} e^{\tfrac{1}{2} \langle \dop y, A \dop y \rangle}=
  \int_{T^{2m}} e^{\sum_{i,j} \dop y_i \wedge \dop y_j B_{i,j-m}} =
  \int_{T^{2m}} \frac{1}{m!} (\sum_{i,j} \dop y_i \wedge \dop y_j B_{i,j-m})^m.
\]
Here the sum is over $i=1, \dots, m$ and $j=m+1, \dots, 2m$. Because
the integral is only nonvanishing when all $\dop y_i$ and $\dop y_j$
occur exactly once, we find
\[
\begin{split}
  \int_{T^{2m}} e^{\frac{1}{2} \langle \dop y, A \dop y \rangle} &=
  \int_{T^{2m}} \sum_{j_1, \dots, j_m} \dop y_1 \wedge \dop y_{j_1}
  \wedge \dots \wedge \dop y_m \wedge  \dop y_{j_m} 
  B_{1,j_1-m} \dots B_{m, j_m-m} \\
  &= \int_{T^{2m}} \sum_{\sigma \in S^m} \sgn(\sigma) \dop y_1 
  \wedge \dop y_{m+1} \wedge \dots \wedge \dop y_m \wedge \dop y_{2m}
  B_{1,\sigma(1)} \dots B_{m,\sigma(m)} \\
  &= \det(B) \vol(T^{2m}) = \sqrt{\det(A)} \vol(T^{2m}).
\end{split}
\]
More general integrals can be computed by `completing the square'.
\begin{equation}
\label{eq:complsqr}
\begin{split}
  \int_{T^{2m}} e^{\frac{1}{2} \langle \dop y, A \dop y \rangle +
  \langle a, \dop y \rangle} &= \int_{T^{2m}} e^{\frac{1}{2} \langle
  \dop y - A^{-1} a, A(y - A^{-1} a) \rangle - 
  \frac{1}{2} \langle A^{-1} a, a \rangle} \\
  &= e^{\frac{1}{2} \langle a, A^{-1} a \rangle} \sqrt{\det(A)}.
\end{split}
\end{equation}
Here $a$ is a vector of 1-forms vanishing on $T^{2m}$, so $\langle
\dop y, a \rangle = - \langle a, \dop y \rangle$. We also use the
antisymmetry of $A$ to rewrite $- \langle A^{-1} a, A \dop y \rangle =
\langle a, \dop y \rangle$. The shift over $A^{-1}a$ does not affect
the Gaussian integral, because we have $2m$ different 1-forms $\dop
y_i - (A^{-1} a)_i$. So wedge products of more than $2m$ factors
vanish automatically. The integral picks out the term with only $\dop
y_i$, because we are in dimension $2m$.

Another simple, but useful formula for integrals of differential forms
on a torus is
\begin{equation}
\label{eq:deltaform}
  \int_{T^m} e^{\langle a, \dop y \rangle} = a_1 \wedge \dots
  \wedge a_m.
\end{equation}
Here $a$ is a vector of 1-forms vanishing on $T^m$.

\begin{raggedright}
\bibliographystyle{hep}
\bibliography{temp,thesis1,thesis2,dbranes,categories}
\end{raggedright}
\end{document}